\documentclass[11pt]{article}
\usepackage{amsmath,amssymb}
\usepackage[dvipdfmx]{graphicx}
\usepackage{graphicx}
\usepackage{url}
\usepackage{braket}

\usepackage{enumerate}

\newcommand{\bs}{\boldsymbol}

\begin{document}
\title{Magnetization without spin : effective Lagrangian  of itinerant electrons.  }    
\author{
{Kenzo Ishikawa} $^{(1,2)}$}
\maketitle
\begin{center}
(1)  Department of Physics, Faculty of Science, \\
Hokkaido University, Sapporo 060-0810, Japan \\
(2)  Natural Science Center,\\
Keio  University, Yokohama 223-8521, Japan \\
\end{center}


\begin{abstract}
Effective Lagrangian  of   itinerant electron system  of  finite density at finite magnetic field  is found to include Chern-Simons term of electromagnetic potentials of  lower scale dimension than those  studied before. This term has an origin in many-body wave function and unique topological  property that   is  independent of  a spin degree of freedom. The coupling strength is proportional to  $\frac{\rho}{eB}$, which  is singular at $B=0$ for a constant charge density. The effective Lagrangian  at a finite $B$ represents physical effects   at  $ B \neq 0$ properly.  A universal  shift of the magnetic field known as  Slater-Pauling curve is derived from the effective Lagrangian.

\end{abstract}
\newpage
 \section{ Effective Lagrangian of vector potentials }

 From Bohr-van Leeuwen theorem, a magnetization does not appear  in a thermodynamic  limit of classical electron  systems.  A quantum correlation  is necessary for matters to show magnetization. A spin and an orbital angular momentum become to satisfy commutation relations and  couple with a magnetic field  through  magnetic moments. This  can lead a magnetization.   The  spin degree of freedom describes effectively  localized  electrons  in insulator. A system of localized electrons  is   described   effectively by  spin Hamiltonian  such as Heisenberg model. The magnetization  are understood   based on a mean field theory.

Conduction electrons in metal are not expressed by   a spin Hamiltonian, but reveals  magnetization. There would be another  mechanism in which  the spin does not play a role.    Magnetization of  metals has a long history and  has not been fully understood yet.   In this paper, a magnetization  of  systems of  charged  particles  is studied with effective Lagrangian of vector potentials. Because the Lagrangian  gives a unified  description of  system's electromagnetic properties, this analysis   may sheds a new light  toward  an itinerant magnetism.

Motion of a free charged particle is along a  straight trajectory in a system of no magnetic field, $B=0$, and that is a circular motion in uniform magnetic field. A trajectory is open in the former and closed and periodic in the latter \cite{taub}. Quantization of charged particles in a  magnetic field   is made through a kinetic energy.  Particle's static energy in a  magnetic field is expressed by a velocity and vector potential as  $e {\vec v} \cdot {\vec A}$, where   $(A_x,A_y)=\frac{B}{2} (-y,x)$ for a constant field in $z$-direction  in a symmetric gauge, $e$ is the unit of electric charge, and  $\vec v$ is a velocity operator. A Larmor radius $R$ for $B \neq 0$ is given using a velocity $v$ and  a magnetic field as $R_B=\frac{m}{eB}v$, where $m$ is the mass. A radius  becomes large for  weak field and for high velocity. Figure 1 shows a radius $R_B$ for  a magnetic field $B$ and a velocity determined from its kinetic energy, $\frac{m_e}{2} v^2 =\frac{3}{2} kT$   for $T=300K$.    
\begin{figure}
\begin{center}
  \includegraphics[scale=.4]{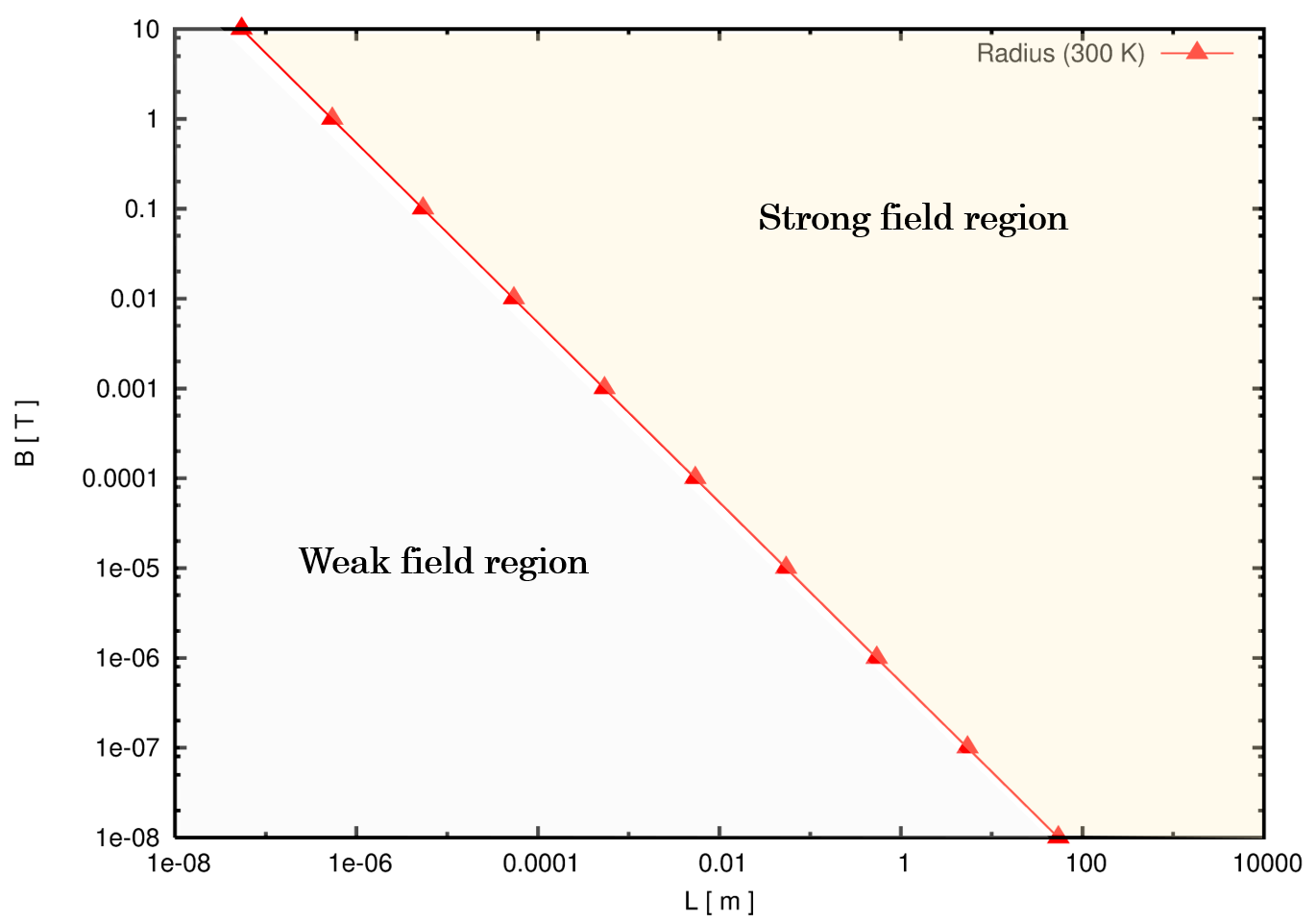}
\caption{ The red line shows the Larmor radius  as a function of the magnetic field.  Magnetic field is in the vertical axis  and the Larmor radius is in the horizontal axis.   In an  area of yellow,  the system is longer than the Larmor radius.   The  temperature  is  $300$  K. 
}
\end{center}
\end{figure}

Quantum effects specific to a magnetic field   are important when a  radius $R_B$ is smaller than a system's size. The electrons in these states   are not represented by plane waves but by those of circular motion of finite radius.  Quantization of closed paths is applied   in the strong field region, and quantization of straight lines is applied in the weak field region.  Physical quantities of these states may involve terms that  are proportional to $\frac{1}{B}$. As this behavior   does not appear  in  a smooth expansion on a series of positive  power $B^n;n=0,1,\cdots$, those quantities in small $B$ are not expressed by correlations of the system at $B=0$. A physical system that reveals non-smooth behavior from  $B=0$ to  $ B \neq 0$ should  be treated by taking into account this term manifestly.   To elucidate $\frac{1}{B}$ term correctly, a quantization  of  periodic orbits at a finite $B$ is necessary.

One-particle states of  charged particles  in a uniform magnetic field  are expressed by Landau levels \cite{Landau-level}. Because these  are orthogonal for
 different energies, their superposition represent isolate states \cite{Ishikawa-Nishio}.  These electrons  are expressed by a field of many components  and reveal various unique features not existing in zero-field systems. 
{ Those features that are specific to a magnetic field are;} (1). One-particle states have degenerate discrete energies, (2) Wave functions are spatially extended of finite dipole moments. (3) A propagator is  classified by a topological invariant \cite{ishikawa-matsuyama}. 
(1) is known to give  a negative magnetic susceptibility. (2) and (3) give an  energy due to macroscopic quantum fluctuations.  These features are manifestly represented in von-Neumann representation for guiding center coordinates, which is expressed  in the following section.

 Using von-Neumann representation,    universal relations between a charged density and a field operator due to  current conservation and commutation relations become simple forms  convenient  to study  quantum corrections.   We  find an effective Lagrangian representing   an adiabatic   shift of  external variables such as magnetic field $B$ or electron density.  A new term of  lower   scale dimension than those studied before of  an origin  in  many-body quantum correlations  in a magnetic field  due to a spectral asymmetry of one-body  Hamiltonian is found. This has no direct connection with a spin degree of freedom, but is expressed by an  induced Chern-Simons term.
The Chern-Simons term   has  topological  origins,  intriguing  properties,  and sizable magnitude.  A coupling strength  in a finite charge density  is very different from the one in a vacuum   and is proportional to $\frac{\rho_e}{B}$, where $\rho_e$ is the electron density.

Integrating an effective Lagrangian over a magnetic field, we find an energy  of the electron system  at a finite $B$. The energy is  proportional to  $\log{B/B_0}$, where $B_0$ is a low field cutoff. A   $B$-dependent energy implies that the itinerant electron system exhibits magnetic properties  without  a spin degree of freedom.    

The present paper is organized in the following manner. Introduction is given in Section 1, and 
 in Section 2  properties of system under discrete transformations and scale transformations,  and field theory with von-Neumann representation for electrons in a magnetic field are studied.  In Section 3,  effective Lagrangian of a system of finite magnetic field is derived  and its implications  to itinerant magnetism  are  analyzed. Conclusion is given in Section 4.

\section{Electrons in a uniform magnetic field}
 An effective action  of the electromagnetic  potentials from a long-distance fluctuation of electrons is obtained. The    action of low scaling dimension which represents responce of the system in a long distance region  emerges   in the sytem of finite magnetic field. This 
term  has been dismissed in previous studies of itinerant magnetism.

\subsection{Symmetry  }
The system has continious  and discontious symmetries.   
 \subsubsection{Space-time inversions }
Gauge invariance and  space inversion  are preserved, but  a time-inversion is  broken in magnetized  system. 
  The  Lagrangian  of electrons and photons in an external electromagnetic field 
 in  a non-relativistic region 
\begin{eqnarray}
\mathcal{L}&=& \psi (x)^{\dagger}  ((p_{0}+e ({A_0^{ext}}(x)+A_{0}(x)) -\frac{1}{2m}(p_{i}+e ({A_i^{ext}}(x)+A_{i}(x) ))^2) \psi (x) \nonumber \\
&+& \mathcal{L}_{em}, \label{lagrangian}
\end{eqnarray}
where $\psi(x)$ is an electron field,  $A_{\mu}(x)$ is an photon field, and  $\mathcal{L}_{em}$ shows the Lagrangian that depends on vector potentials. This   is invariant under the gauge transformation
\begin{eqnarray}
& &A_{\mu}(x) \rightarrow  U^{-1}(x)(A_{\mu}+p_{\mu} ) U(x), \\
& & \psi(x) \rightarrow U(x) \psi(x),
\end{eqnarray}
where $U=e^{i\lambda(x)}$. 
Electromagnetic current  
\begin{eqnarray}
  & & J_0(x)= \psi(x)^{\dagger} \psi(x), \\
& & J_i(x)=  \psi(x)^{\dagger} \frac{1}{m} (p_{i}+e ({A^{ext}}_{i}(x)+A_{i}(x)))   \psi(x) 
\end{eqnarray}
 are  conserved. 
 Ward-Takahashi identity hold in   the system of the external magnetic field.  

The Hamiltonian is given as
\begin{eqnarray}
\mathcal{H}&=& \int dx [\psi (x)^{\dagger}  (-e ({A_0^{ext}}(x)+A_{0}(x)) +\frac{1}{2m}(p_{i}+e ({A_i^{ext}}(x)+A_{i}(x) ))^2) \psi (x)] \nonumber \\
&+& \mathcal{H}_{em}. \label{lagrangian}
\end{eqnarray}
 For a uniform density $A_0^{ext}(x)=constant $ and a uniform magnetic field $A_i^{ext}(x)$,  the vector potential is transformed under a space reflection as,
 \begin{eqnarray}
& & t  \rightarrow  t, \vec x \rightarrow - \vec x \\
& & \vec A(x) \rightarrow - \vec A(x) , \vec A(x)^{ext} \rightarrow - \vec A(x)^{ext}, 
\end{eqnarray}
and is transformed under a time inversion as,
 \begin{eqnarray}
& & t  \rightarrow - t, \vec x \rightarrow  \vec x \\
& & \vec A(x) \rightarrow - \vec A(x) , \vec A(x)^{ext} \rightarrow  \vec A(x)^{ext} . 
\end{eqnarray}
The system is invariant under  a space inversion, but is not invariant under  a time-inversion. 
\subsubsection{Scale transformation}
Operators  are transformed under a scale transformation  according to own scale dimensions. 
 $D$-spatial dimensions is studied. 

 Under scale transformation in spatial coordinates
\begin{eqnarray}
x_i \rightarrow \lambda x_i, \label{scale-transformation}
\end{eqnarray}
fields are transformed in accord with commutation relations,
\begin{eqnarray}
& &[\psi(\vec x,x_0),\psi^{\dagger}(\vec y,y_0)]_{+} \delta(x_0-y_0)= i\hbar \delta^{(D+1)}(x-y), \\
& &[A_i(\vec x,x_0),\dot A_j(\vec y,y_0)]_{-} \delta(x_0-y_0)= i\hbar \delta^{(D+1)}(x-y), 
\end{eqnarray} 
as 
\begin{eqnarray}
& &\psi(\lambda x_i)= \lambda^{-D/2} \psi(x_i) ,\\
& &A_i( \lambda x_i)=\lambda^{-(D-1)/2} A_i(x_i).
\end{eqnarray} 
Then composite operators are transformed as
\begin{eqnarray}
& &J_0(\lambda x)=\psi^{\dagger}( \lambda x_i)\psi(\lambda x_i)= \lambda^{-D} \psi(x_i)^{\dagger} \psi(x_i), \label{charge_d}\\
& &\psi^{\dagger}( \lambda x_i)\frac{\vec p}{m \lambda}\psi(\lambda x_i)= \lambda^{-D-1}\psi^{\dagger}(  x_i)\frac{\vec p}{m}\psi( x_i), \label{current_d}\\ 
& &\psi^{\dagger}( \lambda x_i)\frac{(\vec p+e\vec A)_{r}}{m}\psi(\lambda x_i)= \lambda^{-D}\psi^{\dagger}(  x_i)\frac{(\vec p+e\vec A)_{r}}{m}\psi( x_i) \label{current_d_m},
\end{eqnarray} 
where a velocity $\frac{\vec p}{m}$ has a dimension $\lambda^{-1} $ in Eq.$(\ref{current_d} )$, whereas a velocity  in a magnetic field $(\vec p+e\vec A)_{r} $ has a dimension $\lambda^{0} $ due to  a circular  motion in a uniform magnetic field. Details  are  shown in a next  section. 

 Dimensions of products of electromagnetic current and  potentials are 
\begin{eqnarray}
& &J_0(\lambda x) A_0(\lambda x)= \lambda^{-\frac{3D}{2}+\frac{1}{2}}  J_0( x) A_0(x), \\
& &J_i(\lambda x) A_i(\lambda x)= \lambda^{-\frac{3D}{2}-\frac{1}{2}}  J_i( x) A_i(x), \\ 
& &J_i^{mag}(\lambda x) A_i(\lambda x)= \lambda^{-\frac{3D}{2}+\frac{1}{2}}  J_i^{mag}( x) A_i(x).  
\end{eqnarray}

Spin dependent  energy  proportional to  the magnetic field has a dimension
\begin{eqnarray}
& &J_0(\lambda x)^{spin} B(\lambda x)= \lambda^{-\frac{3D}{2}-\frac{1}{2}}  J_0( x)^{spin} B(x), \\
& &J_0^{spin}( x) =\sum_s \psi_s^{\dagger} s_z \psi_s.   
\end{eqnarray}
 
Dimensions of space integrals of the above products  are 
\begin{eqnarray}
& &\int d^Dx J_0( x) A_0( x) \rightarrow  \lambda^{-\frac{D}{2}+\frac{1}{2}} \int d^D x J_0( x) A_0(x), \label{elemag_1}\\
& &\int d^D x J_i( x) A_i( x) \rightarrow \lambda^{-\frac{D}{2}-\frac{1}{2}} \int d^Dx  J_i( x) A_i(x), \label{elemag_2}\\ 
& &\int d^D x J_i^{mag}( x) A_i( x) \rightarrow \lambda^{-\frac{D}{2}+\frac{1}{2}} \int d^D x J_i^{mag}( x) A_i(x). \label{elemag_3} 
\end{eqnarray}
and that of space integral of spin dependent  energy  proportional to  the magnetic field is
\begin{eqnarray}
& &\int d^D x J_0( x)^{spin} B( x) \rightarrow  \lambda^{-\frac{D}{2}-\frac{1}{2}} \int d^Dx  J_0( x)^{spin} B(x), 
\label{spin_1} 
\end{eqnarray}
   
Dimensions of products of vector potentials are
\begin{eqnarray}
& &\frac{1}{\lambda}\partial_l A_i ( \lambda x_i) \frac{1}{\lambda } \partial_k A_i(\lambda x_i)=\lambda^{-D-1} \partial_l A_i(x_i) \partial_kA_i(x_i),\\
& &A_i( \lambda x_i)\frac{\partial}{ \lambda \partial x_k}A_i(\lambda x_i)=\lambda^{-D} A_i(x_i) \frac{\partial}{\partial x_k}A_i(x_i).
\end{eqnarray} 
Dimensions of integrals of products of vector potentials are
\begin{eqnarray}
& &\int d^D x \partial_l A_i (  x_i) { \partial_k }A_i(\lambda x_i) \rightarrow\lambda^{-1} \int d^D x\partial_l A_i(x_i) \partial_kA_i(x_i),\\
& &\int d^D xA_i(  x_i){  \partial_k}A_i( x_i) \rightarrow \lambda^{0} \int d^Dx A_i(x_i) \frac{\partial}{\partial x_k}A_i(x_i).
\end{eqnarray} 

\subsubsection{Scale dimension}
Products of momenta  are transformed according to dimensions under a scale transformation to
\begin{eqnarray}
& &p_l \rightarrow \frac{1}{\lambda} p_l, \nonumber \\ 
& &p_{l_1}p_{l_2} \rightarrow \frac{1}{\lambda^2} p_{l_1}p_{l_2} \nonumber \\
& &p_{l_1} \cdots p_{l_M} \rightarrow \frac{1}{\lambda^M} p_{l_1}\cdots p_{l_M}\nonumber 
\end{eqnarray}
A magnitude of a product of M momenta in a small momentum region  of $p_l =O(\epsilon)$
is ${\epsilon}^M$.   Physical phenomena of large scale are described by amplitude or Green's function of small  momenta of $p_l =O(\epsilon)$. Quantities become more suppressed and  become small in low energy regions, i.e., long distance region,  according to scale dimensions.
       
Spin interaction energy Eq.$(\ref{spin_1})$ and an electromagnetic interaction Eq.$(\ref{elemag_2})$ are important in a short distance region, while   charge interaction energies Eqs.$(\ref{elemag_1}), (\ref{elemag_3})$ are important in a long distance region. 

\subsubsection{Electron correlations }
(I) Coulomb interaction works  universally between charge densities.

An  interaction energy between charges  is transformed 
 to
\begin{eqnarray}
& &\int d^D x d^D y J_0(\vec x) V(x-y) J_0(\vec y) \approx   \int d^D x  J_0(\vec x) \tilde V_0 J_0(\vec x)  \rightarrow  \lambda^{-3}\int d^D x  J_0(\vec x) \tilde V_0 J_0(\vec x) \nonumber \\
& &
\end{eqnarray}
 for a screened   Coulomb interaction
\begin{eqnarray}
V(x-y)=\frac{V_0}{|\vec x-\vec y|} e^{-\mu|\vec x -\vec y|},
\end{eqnarray}
where
\begin{eqnarray}
\tilde V_0=\int d {\vec y}\frac{V_0}{|\vec x-\vec y|} e^{-\mu|\vec x -\vec y|}.
\end{eqnarray}

(II) Electromagnetic energies of spin correlations.

As a  spin interacts with a magnetic field, $B$ dependent energy  is  expressed as  
\begin{eqnarray}
& &\int d^D x  B(\vec x) \langle J_s(\vec x) \rangle  \rightarrow \lambda^{D-2}  \int d^D x  B(\vec x) \langle J_s(x) \rangle 
\end{eqnarray}
and 
\begin{eqnarray}
& &\int d^D x d^D y B(\vec x) V_s(x-y) B(\vec y) \rightarrow \lambda^{-1}  \int d^D x  B(\vec x) V_s^0 B(\vec x)  
\end{eqnarray}
for a short  range spin correlation
\begin{eqnarray}
\int d^D y V_s(x-y)=\int d^Dx \langle J_s(x) J_s(y) \rangle= V_s^0.
\end{eqnarray}
Here 
\begin{eqnarray}
& &\langle J_s(\vec x) \rangle \\
& &\langle J_s(x) J_s(y) \rangle
\end{eqnarray}
are correlation functions of spin current operator.  Correlations of  spin has been studied widely in  literatures.   

(III) Electromagnetic energies due to  electron's current correlations.
From current correlations, $A_{\mu}$ dependent energy  is  expressed as  
\begin{eqnarray}
& &\int d^D x  A_{\mu}(\vec x) \langle J^{\mu}(\vec x) \rangle =0  
\end{eqnarray}
and 
\begin{eqnarray}
& &\int d^D x d^D y A_{\mu}(\vec x) \Pi^{\mu,\nu}(x-y) A_{\nu}(\vec y) \rightarrow \lambda^{1+\lambda_{\pi}}  \int d^D x   A_{\mu}(\vec x) \Pi^{\mu,\nu}(x-y) A_{\nu}(\vec y) \nonumber \label{two_vector} \\
& &    
\end{eqnarray}
Here  current correlation functions satisfy 
\begin{eqnarray}
& &\langle J_{\mu}(\vec x) \rangle =0, \\
& &\int dy \langle J_{\mu}(x) J_{\nu}(y) \rangle \rightarrow \lambda^{\lambda_\pi} \int dy \langle J_{\mu}(x) J_{\nu}(y) \rangle,
\end{eqnarray}
where $\lambda_{\pi}$ is a scale dimension of two current correlation functions     computed later. We will see that $\lambda_{\pi}=-1$, and  Eq. $(\ref{two_vector})$  gives a dominant contribution in a long distance region.  
  \subsection{ Finite  ${B}$ expansion }     
A current operator Eq.$(\ref{current_d_m})$ in a system of magnetic field has a larger scale dimension than the one of no magnetic field,  and a magnitude of its expectation value  in  long distance region becomes enhanced over current Eq.$(\ref{current_d})$. A representation of the electrons that reveals this behavior manifestly is von Neumann representation of  electrons  in Landau levels. Computation of  an effective Lagrangian of vector potentials can be  made in a transparent manner.     

A  system of  finite  electron density and magnetic field  parallel to  $z$-direction  is expressed  by  $(A^\text{ext}_x,A^\text{ext}_y)=\frac{B}{2}(-y,x)$ and  $A^{\text{ext}}_{0}(x)=\frac{\mu_e}{e}$. In this system, electron states of negative energies and positive energies below the Fermi energy,  $ 0 \leq E \leq \mu_e$, are occupied. The effective Lagrangian from  the former ones is the one of the vacuum and was obtained  in   \cite{erdas}.  The latter one with Pauli-Villars regularization was obtained in \cite{sakita}. Universal properties of the Lagrangian are clarified and applied to itinerant magnetism  in the present paper.  

\subsubsection{von Neumann lattice representation:  $H_0=\frac{p^2}{2m},D=2$ }

We study  a system of  low energy electrons which are expressed by energies quadratic in momentum in two spatial dimensions for simplicity.  It is straightforward to extend to  high energy  using the  relativistic Landau levels, which are given in the appendix, and to a system of $D=3$. The single electron Hamiltonian  is
\begin{align}
{H}_0=  \frac{({p_{x}}+e{ A}_x^\text{ext})^2}{  2m}+\frac{({ p_{y}}+e{ A}_y^\text{ext})^2}{  2m}+\sigma_z g \frac{eB}{2m} 
\end{align}
 To elucidate higher order corrections  we decompose  the electron field 
 in terms   of the  Landau levels. A spin degree is ignored for a while. 

As the electron's motion is circular around an arbitrary  center, it is transparent to express the electron field  with center coordinates $(X,Y)$ and relative coordinates $(\xi,\eta)$. One-body Hamiltonian in $(x,y)$ space is described by the relative coordinates $(\xi,\eta)$,   and commutes with  center coordinates $(X,Y)$ as,   
\begin{eqnarray}
& &\xi =(eB)^{-1}(p_x+eA_x^\text{ext}), \eta =(eB)^{-1}(p_y+eA_y^\text{ext}),~~X=x-\xi, Y=y- \eta, \nonumber \\
& & \\
& &[ \xi,\eta ]=-[X,Y]=\frac{a^2}{2 \pi i}, a=\sqrt{\frac{2 \pi \hbar}{eB}},\\
& &{H^0}_{(xy)} = \frac{1}{2} m {\omega_c}^2 (\xi^2+\eta^2)+\frac{\sigma_z}{2}\omega_c, \omega_c=\frac{eB}{m}.
\end{eqnarray}
 The  eigenstates of this    Hamiltonian and those of the center coordinates,  
 \begin{eqnarray}
& &{H^0}_{(xy)}|l \rangle = E_l| l \rangle,E_l= (l+\frac{1}{2} \pm\frac{1}{2}) \hbar \omega_c,   l=0,1,2 \cdots   \\
& &(X+ iY) |\alpha_{mn} \rangle =z_{mn} | \alpha_{mn} \rangle ,z_{mn}= (m \omega_x+i n {\omega_y} ) \nonumber \\
& &|\alpha_{mn} \rangle= e^{i \pi(m+n+mn)+\sqrt{\pi} (A^{\dagger}\frac{z_{mn}}{a}-A\frac{z_{mn}^{*}}{a})} |\alpha_{00} \rangle , \\
& &A=\frac{\sqrt \pi}{a}(X+iY), [A,A^{\dagger}]=1.
\end{eqnarray}
where complex numbers $\omega_x$ and $\omega_y$  satisfy 
$\text{Im} [\omega_x^{*} \omega_y]=1$.
 Many-body  quantum  fluctuations are  studied easily with the  momentum states of the guiding centers
\begin{eqnarray}
& &|\alpha_{\bs p} \rangle= \sum_{mn}e^{ip_x m+ip_y n} |\alpha_{mn} \rangle , \langle \alpha_{\bs p}|\alpha_{\bs p'} \rangle=\alpha({\bs p}) (2\pi)^2 \delta({\bs p}-{\bs p}'-2 \pi  \bs N) \nonumber \\
& &\\
& &| \beta_{\bs p} \rangle= \frac{1}{\beta(\bs p)} |\alpha_{\bs p} \rangle, \alpha( \bs p)= \beta^{*}( \bs p)  \beta( \bs p),
\end{eqnarray}
where $\bs N$ is a two dimensional vector of  integer components, $\bs p$ is a two dimensional momentum in a magnetic Brillouin  zone, and   $\alpha(\bs p)$ is expressed by Jacobi $\theta $ -function. This basis,  $|l, \beta_{\bs p} \rangle$, the von Neumann basis,   is convenient to investigate   a many-body dynamics of   quantum Hall systems \cite{imos,iaim}, and neutrinos in the solar corona  \cite{corona}.

For a completeness of paper, we briefly review  hereafter   an expansion of the electron field  with the  von Neumann basis,
\begin{eqnarray}
& &\psi({\vec x},t)= \int_{BZ}\frac{d^2 p}{ 
 (2\pi)^2}\sum_{l=0}^{\infty} b_l({\bs p},t) \langle {\bs
 x}|l,\beta_{\bs p}\rangle,\ E_l(p_z)=E_l, \\
& &\langle {\vec x}| l,\beta_{\bs p} ,  \rangle = \langle {\bs x}| l,\beta_{\bs p}  \rangle  \nonumber 
\end{eqnarray}
where $E_l$ is the energy of Landau level and the BZ stands for the magnetic Brillouin zone  
\begin{eqnarray}
p_i \leq \frac{\pi}{ d},\ d=\sqrt \frac{2\pi \hbar}{  {eB}}.
\end{eqnarray} 
Creation and annihilation operators  satisfy
\begin{eqnarray}
& &\{ b_l({\bf  p}\,), b_{l'}^{\dagger}({\bf p}') \}= \delta_{l l'}(2 \pi)^2 ({\bs p}-{\bs p}\,'-2\pi{\bs N})
 e^{i\phi({\bs p},{\bs N})},\\
& &\phi({\bs p},{\bs N})= \pi(N_x+N_y)-N_yp_x,\ N_i: \text {integer}  \nonumber.
\end{eqnarray}
The one body Hamiltonian  is diagonal but the  density and current in the $(t,x,y,z)$ space  are expressed as,
\begin{eqnarray}
j_0({\bf k})
 &=&\sum_{l,l'}  \int_{BZ}\frac{d^2p}{
 (2\pi)^2} { b}_l^{\dagger}({\bf p}) \Gamma_0({\bf k})_{l,l'} b_{l'}({\bf p}+a\hat k)   \label{form-factor_1}\\
{\bs j}({\bf  k})
 &=& \sum_{l,l'}  \int_{BZ}\frac{d^2p}{
 (2\pi)^2}  b_l^{\dagger}({\bf p}){\bs \Gamma({\bf k})}_{l,l'} b_{l'}({\bf p}+a\hat k) \label{form-factor_2} 
\end{eqnarray}
where $\hat {\bs k}$ is dual momentum of ${\bs k}$
\begin{eqnarray}
& &\hat k_i=W_{ij} k_j,
\end{eqnarray}
\[
W=
\left(
\begin{array}{cc}
\text{Re} [ {\omega}_x] & \text{Im} [ {\omega}_x]  \\
\text{Re} [ {\omega}_y] & \text{Im} [ {\omega}_y]
\end{array}
\right) 
\]
and  
\begin{eqnarray}
\Gamma_0({\bf k})_{l,l'} =\langle f_l| e^{-i{\bs k}\cdot{\bs 
 \zeta}}  | f_{l'}  \rangle,  {\bs \Gamma({\bf k})}_{l,l'}= \langle f_l| \frac{1}{ 2}\{{\bs v}, e^{-i{\bs k}\cdot{\bs
 \zeta}} \} | f_{l'} \rangle, 
 \end{eqnarray}
where $\bs v=\frac{\bs p+e\bs A_{ext}}{m}$ and   $\bs \zeta=(\xi,\eta)$. The form factor $\Gamma_0(\bs k)$  in
 Eq.$(\ref{form-factor_1} )$ and $\bs \Gamma(\bs k)$  in
 Eq.$(\ref{form-factor_2} )$ is expressed by a polynomials in $\bs k$. The constant term, $\Gamma_0(0)$ is proportional to the electron charge and  the linear  term $\frac{\partial }{\partial k_i} \Gamma(0)$ is proportional to the dipole moment, which has non-vanishing matrix element  between nearest neighbor  Landau levels, i.e., $l'=l \pm 1$.

A  transformed  electron operators with a unitary matrix 
\begin{eqnarray}
& &\hat b_l(\bs p)=\sum_{l'}U_{l l'}(\bs p) b_{l'}(\bs p)\\
& &U_{l l'}(\bs p)=\langle f_l | e^{i{\bs p} \cdot {\bs {\hat  \xi}}/a-\frac{ i}{4\pi} p_x p_y}|f_{l'} \rangle, \hat \xi_i=W_{ij} \xi_j. 
\end{eqnarray}   
 satisfy  
 \begin{eqnarray}
 & &\{ {\hat b}_l({\bf p}\,), {\hat b}_{l'}^{\dagger}({\bf  p}') \}= \delta_{l l'}(2 \pi)^2 ({\bs p}-{\bs p}\,'-2\pi{\bs N}) \Lambda_{l l'}(\bs N),\nonumber \\
& & \\
& &{\hat b}_l({\bs p}+2 \pi {\bs N})=  \Lambda_{l l'}(\bs N) {\hat b}_{l'}({\bs p}) \Lambda_{l l'}(\bs N)=(-1)^{N_x+N_y} U_{l l'}(2 \pi \bs N)  \nonumber,
 \end{eqnarray}
and lead   a simple expression of  the charge and current operators 
 \begin{eqnarray}
& &j_0({\bf k})= \sum_{l}  \int_{BZ}\frac{d^2p}{
 (2\pi)^2} {\hat b}_l^{\dagger}({\bf p})  {\hat b}_{l}({\bf
 p}+a\hat k)   \label{form-factor_{20}} \\
& &{\bs j}({\bf k})= \sum_{l,l'} \int_{BZ}\frac{d^2p}{
 (2\pi)^2}  {\hat b}_l^{\dagger}({\bf p}) \{ \bs v+\frac{a \omega_c}{2\pi}(W^{-1}\bs p+\frac{a}{2} \bs k)\}_{l,l'} {\hat b}_{l'}({\bf  p}+a\hat k). 
\end{eqnarray}
A form factor dose not appear in a matrix element of the charge operator, accordingly a commutation relation between the charge density and the electron operators 
becomes straightforward.  
\begin{figure}[h]
\begin{center}
\includegraphics{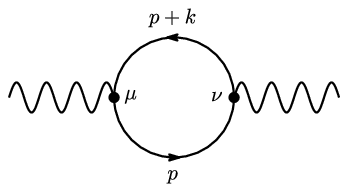}
\caption{The diagram of the current correlation function is shown. The real line represents the electrons and the wave line represents the photon or the neutrino current.}
\label{fig:selfenergy}
\end{center}
\end{figure}
The electron's propagator is defined as a matrix in the magnetic Brillouin zone and Landau level index,
\begin{eqnarray}
\langle 0|{{\hat b}_m(\bs p) {\hat b}_n^{\dagger} (\bs p') }|0\rangle=(2 \pi)^3 \delta(\bs p-\bs p')  {\hat S}_{mn}(\bs p),
\end{eqnarray} 
and a vertex part is written as 
\begin{eqnarray}
\langle j^{\mu}(q) {\hat b}_m(p) {\hat b}_{n}(p') \rangle= (2\pi)^3 \delta(p+Q-p') {\hat S}_{mm'}(p,p+Q) {\hat \Gamma}^{\mu}_{m',n'} (p,p+Q) {\hat S}_{n'n} (p+Q)  
\end{eqnarray}
where $Q^{\mu}=(q_0,a \hat q_x,a \hat q_y)=t^{\mu}_{\nu} q^{\nu}$.
The vertex part  satisfies   a magnetic Ward-Takahashi identity, 
\begin{eqnarray}
{\hat \Gamma}_{\mu}(p,p)=t_{\mu}^{\nu} \frac{\partial {{\hat S}(p)}^{-1}}{\partial p^{\nu}}, \label{W-T identity} 
\end{eqnarray}
and 
\begin{eqnarray}
& &\frac{\partial }{\partial Q_{\nu}} {\hat \Gamma}_{\mu} (p,p+Q)|_{Q=0} \nonumber \\ 
& &=
\begin{cases}
0; \nu=0,3 ~ ; \mu=0,3 \\
\delta_{ij} ; \nu=i,\mu=j,  i,j=1,2
\end{cases} \label{vertex_d}
\end{eqnarray}
\subsubsection{von Neumann lattice representation:tight binding model}
In a tight-binding model, Hamiltonian is 
\begin{eqnarray}
H_0=K({e^{ipa}-e^{-ipa}}),
\end{eqnarray}

Spectrum at a weak magnetic field where (1) a  magnetic flux per plaquette  $\Phi$ is much smaller than a unit of flux $\Phi_0$, $\Phi <<\Phi_0$, and (2)  $\Phi \approx  \Phi_0$ are very different. (1) is a weak magnetic field and (2) is a strong magnetic field. 

von Neumann representation and Hofstadter butterfly Peierls  substitution

\subsection{Effective action }

An action including the  fluctuations of the effective electromagnetic field in Eq.$(\ref{lagrangian})$
 $ A_{\mu}$ is  
\begin{eqnarray}
& &S=S_0+ \int d^3x [  e j_0(x) { A}_0(x) - e {\vec j}(x)\cdot{\vec { A}}(x) +O({ A}_{\mu}^2)] \label{action} ,\\
& &S_0= \int dt  \int_{BZ}\frac{d^2 p }{  (2\pi)^2} 
 b_l^{\dagger}({\bf p}\,) 
\left[i\hbar \frac{\partial}{
 \partial t} -E_l
\right]
 b_l ({\bf  p})\nonumber
\end{eqnarray}
where the current operators $j_0(x)$ and  ${\vec j}(x)$ are given below. 
Integrating over the electron fields in path integral method of a system described by Eq.$(\ref{action})$ or applying a standard perturbative expansions of the transition amplitude, we obtain  the effective Lagrangian for the  photon by applying the derivative expansion on the fields $ A_{\mu}(x)  $  as   
\begin{eqnarray}
S_{eff}=\frac{e^2}{ 2} \int d^3x d^3y  A_{\mu}(x) \pi^{\mu\nu}(x-y)
 A_{\nu}(y),   \label{effective-action}
\end{eqnarray}
where 
the current correlation 
function $\pi^{\mu\nu}(x-y)$ is expressed in Fig.$(\ref{fig:selfenergy})$  and its Fourier transform is given as
\begin{align}
\pi_{\mu\nu}({\vec k})= \int \frac{d^3 p}{ (2\pi)^4}  \text{Tr}[{\hat \Gamma}_{\mu}(p,p+k){\hat S}(p+k){\hat \Gamma}_{\nu}(p+k,p)
{\hat  S}(p) ],
\end{align}
where the sum in $Tr $ is made in  the Landau level index. 
Expansion coefficients of   $ \pi_{\mu\nu}({\vec k})$ in a power series on ${\vec k}$,  gives the strengths of the effective interaction. Among the coefficients, the linear term is most important   in the long distance region, and its  coefficient, i.e.,  the slope at the origin, is  finite now  for the space $ (\mu,\nu, \rho=0,x,y) $ due to the inversion symmetry breaking. This  is  expressed as a topological number of the propagator, using the Ward-Takahashi identity Eq.$( \ref{W-T identity})$,  
\begin{align}
&\frac{\partial}{ \partial k_{\rho}}\pi_{\mu\nu}(k)|_{k=0}=
 \epsilon_{\mu\nu\rho} \frac{1}{ 2\pi}N_{w} , \\
& N_w=\frac{ \epsilon^{\mu\nu\rho}}{3!} 
\int \frac{d^3 p}{ (2\pi)^3} \text{Tr}\left[
{\hat S}(p) \frac{\partial {\hat S(p)}^{-1}}{ \partial
 p_{\mu} }{\hat  S}(p) \frac{\partial {\hat S(p)}^{-1}}{ \partial
  p_{\nu}} {\hat  S}(p) \frac{\partial {\hat S(p)}^{-1}}{ \partial
  p_{\rho}} \right] \label{topological_n}.
\end{align}
 $N_w$ is the winding number of the propagator ${\hat S}(p) $.  Its value, which depends on the magnetic field, is evaluated easily by   substituting  
\begin{eqnarray}
\tilde S(\bs p)=U^{\dagger}(\bs p) \tilde S(\bs p) U(\bs p),
 \end{eqnarray} 
into Eq.$(\ref{topological_n})$.
$N_w$ counts the number of Landau bands below the Fermi energy,
\begin{eqnarray}
& &N_w=\frac{\rho}{\rho_0}, \rho_0=\frac{eB}{ 2\pi \hbar},
\end{eqnarray} 
where $\rho$ is the electron density and  $\rho_0$ is  the density of the Landau levels. 
This form is equivalent to the quantum Hall effect \cite{ishikawa} \cite{Coleman-Hill}. 
This topological  invariant  is mathematically   similar  to TKNQ integer  \cite{TK}, but represents physically different quantities. The former is proportional to a filling fraction of Landau levels, whereas the latter is not.

\section{Chern-Simons form at a finite $B$}
The effective action for small change of $A_{\mu}(x)$ is expressed by   Eq.$(\ref{effective-action})$, and  is decomposed to  the lowest dimensional
 term and higher dimensional terms, 
\begin{eqnarray}
& &S_{eff}=S_\text{CS}+S_{high}, \\
& & S_\text{CS}=e^2\frac{\sigma_{xy}^{(4)}}{ 2} \int d^3 x 
 \epsilon^{\mu\nu\rho} { A}_{\mu}
 \partial_{\nu}{ A}_{\rho}, \label{chern-simon} \\
& &\sigma_{xy}^{(4)}=\frac{1}{ 2\pi} \nu^{(4)},\ \nu^{(4)}={ 2\pi} \nu
= \frac{2\pi n_e}{  eB },
\end{eqnarray}
where $S_{high}$ is the higher dimensional term, $n_e$ is the electron density, in the natural unit of  $c=\hbar=1$.
The Lagrangian in $S_{CS}$ is proportional to a momentum and gives  the dominant contribution in the long distance region, which vanishes in a system of zero-magnetic field. 
This term is invariant under a  gauge transformation, and violates time reversal invariance.

The effective Lagrangian in $S_{CS}$, the Chern-Simons (CS) term \cite{imos,iaim,ishikawa},  is
\begin{eqnarray}
\mathcal L_\text{int}= e^2\frac{1}{2}\frac{\nu^{(4)}}{2\pi}\epsilon^{\alpha \beta \gamma} 
 A_{\alpha} \partial_{\beta}  A_{\gamma}  +O({
 F_{\alpha\beta}}^2);\ \alpha,\beta,\gamma=(0,1,2)
\label{effective_lagrangian},
\end{eqnarray}   
where the magnetic field is along the 3rd axis, and  $  F_{\alpha\beta} =\partial_{\alpha}
A_{\beta}-\partial_{\beta}  A_{\alpha}$. 
\subsection{Implications of Chern-Simons term } 
From the effective Lagrangian Eq.$(\ref{effective_lagrangian})$, expectation values of a current and a density is computed.  
\subsubsection{ Hall effects}
The effective action Eq.$(\ref{effective_lagrangian})$ represents  a response  of  the system's energy  under a small variation  of $A_0$  and $A_i(x)$. A derivative of the Lagrangian density with respect to the vector potential,
\begin{eqnarray}
J_{i}(x)=\frac{\delta  \mathcal L_{\text{int}}}{\delta A_{i}(x)} 
\end{eqnarray}
 is a current in an $i$ direction. Substituting Eq.$(\ref{effective_lagrangian} )$, we have
a current proportional to an electric field  
\begin{eqnarray}
J_{i}(x)=  \sigma_H \epsilon^{i \beta \gamma} 
 \partial_{\beta}  A_{\gamma},  
\end{eqnarray}
where a conductance is given by 
\begin{eqnarray}
\sigma_H=e^2\frac{1}{2}\frac{\nu^{(4)}}{2\pi}.\label{hall_conductance}
\end{eqnarray} 
From an anti-symmetric property of $\epsilon^{i \beta \gamma}$, the current is perpendicular to  an electric field, and $\sigma_H$  represents Hall conductance. The effective Lagrangian density leads the  Hall effect.    
\subsubsection{ Varying chemical potential and magnetic field } 
Eq.$(\ref{effective_lagrangian})$ shows another physical effect. A physical system where a   magnetic field $\delta B$ and a  chemical potential $\delta E_F$ are added  in the strong field region in Fig.1,   vector potentials and a scalar potential  vary as, 
\begin{eqnarray}
\partial_1 A_2(x)-\partial_2 A_1(x)= \delta B, A_0=\delta E_F.
\end{eqnarray}
An effective Lagrangian density  under these variations is obtained by substituting  these variables into  Eq.$(\ref{effective_lagrangian})$, as 
\begin{eqnarray}
\mathcal L_\text{int}= e \frac{1}{2}\frac{1 }{2\pi}\frac{2\pi n_e}{  eB }
 (\delta E_F ) \delta B  +O({\delta B}^2);\ \alpha,\beta,\gamma=(0,1,2)
\label{effective-magnetic field lagrangian}.
\end{eqnarray} 
As the effective Lagrangian Eq.$(\ref{effective-magnetic field lagrangian} )$ is linear in a variation  of magnetic field $\delta B$,    a stationary magnetic field makes a shift when a density is varied.   This term has an origin in electrons in Landau levels and  is independent of  a spin degree of freedom. 

Effective Lagrangian for a finite $B$ and a finite density is obtained by integrating the effective Lagrangian Eq.$( \ref{effective-magnetic field lagrangian})$ over $B$ and $n_e$. For a case $B >0$ and $\delta B >0$, we have
 \begin{eqnarray}
L_{int}&=& \int_{B_0}^{B} d B' \int_{0}^{N_e}  d E_F e \frac{1}{2}\frac{1 }{2\pi}\frac{2\pi n_e}{ | e|B' }
   = - \frac{1}{2}  \int_0^{N_e}  { n_e}  d E_F  \int_{B_0}^{B} \frac{dB' }{  B' } \nonumber \\
&=& - \frac{1}{2} \int_0^{N_e} { n_e}   d E_F  \log (\frac{B}{B_0}), \label{induced_lagrangian} 
\end{eqnarray}
where $B_0$ is a lower cutoff, which depends on a size of the system and is studied later.
A coefficient 
\begin{eqnarray}
\int_0^{N_e} { n_e}   d E_F 
\end{eqnarray}
depends on $n_e$. In a simple situation of 
\begin{eqnarray}
n_e= \rho_0  E_F,
\end{eqnarray} 
\begin{eqnarray}
\int_0^{N_e} { n_e}   d E_F =\frac{1}{2\rho_0} N_e^2.
\end{eqnarray}
Then we have the effective Lagrangian
\begin{eqnarray}
L_{int}&=& - \frac{1}{2\rho_0} N_e^2   \log (\frac{B}{B_0}). \label{induced_lagrangian_t} 
\end{eqnarray}
The action is given by an integral over the space time 
\begin{eqnarray}
S_{int}&=& - \int d^3 x \frac{1}{2\rho_0} N_e^2  \log (\frac{B}{B_0}). \label{induced_action_t} 
\end{eqnarray}
\subsection{Combining  short-distance spin fluctuation  }
 A contribution to a free energy from  a short range correlation due to   spin fluctuations has been widely studied in literature. Here we assume various forms for a free energy density and study effects of additional Chern-Simons terms,
 \begin{eqnarray}
F= F_{short}   - \frac{1}{2} \int_0^{N_e} { n_e}  \delta E_F  \log (\frac{B}{B_0}).
\end{eqnarray} 

\subsubsection{ Short-distance magnetization }

 For a case that $F_{short}$ has a minimum at a finite $B_{0,s}$, 
 a  free energy is  given by 
\begin{eqnarray}
F= \frac{c}{2} (B-B_{0,s})^2    - \frac{1}{2} \int_0^{N_e} { n_e}  \delta E_F  \log (\frac{B}{B_0}).
\end{eqnarray}
Derivative of $F$ with respect to $B$ is computed  as
\begin{eqnarray}
& &\frac{ \partial F}{\partial B}= c(B-B_{0,s})-\frac{1}{2} \int_0^{N_e} { n_e} \delta E_F  \frac{1}{B}. 
\end{eqnarray}
A   stationary magnetic field is solved from   
\begin{eqnarray}
& &cB(B-B_{0,s})-\frac{1}{2} \int_0^{N_e} { n_e}  \delta E_F=0 
\end{eqnarray}
as
\begin{eqnarray}
& &B= \frac{cB_0\pm\sqrt{c^2B_{0,s}^2+2c\int_0^{N_e} n_e \delta E_F}}{2c}.  
\end{eqnarray}
For  small $\delta B=B-B_{o.s}$ and $ \delta E_F$, it follows that
\begin{eqnarray}
cB_{0,s} \delta B=\frac{1} {2}  { N_e}  \delta E_F,
\end{eqnarray}
and 
\begin{eqnarray}
 \delta B=\frac{1} {2cB_{0.s}}  { N_e}  \delta E_F.
\end{eqnarray}
 A shift of the magnetic field, $\delta B$, is proportional to the electron density and a shift of the Fermi energy. Accordingly, a shift per an electron is given by
\begin{eqnarray}
 \frac{\delta B}{N_e}= \frac{1} {2cB_{0.s}}  \delta E_F, \label{slator-pauling}
\end{eqnarray}
and  is proportional to $\delta E_F$.   A fact that the proportional constant depends only on $B_{0_s}$ is a feature of  Slater-Pauling curves \cite{kubler}.
\subsubsection{Short-distance non-magnetization}
For $B_{0,s}=0$,
\begin{eqnarray}
F= \frac{c}{2} B^2    - \frac{1}{2} \int_0^{N_e} { n_e}  \delta E_F  \log (\frac{B}{B_0}). \label{positive_B}
\end{eqnarray}
Derivative of $F$ with respect to $B$ is computed  as
\begin{eqnarray}
& &\frac{d F}{dB}=cB-\frac{1}{2B}n_e \delta E_F =0 \\
& &B^2=\frac{1}{2c}n_e \delta E_F 
\end{eqnarray}
A solution is 
\begin{eqnarray}
B=  \sqrt{\frac{1}{2c}n_e \delta E_F }.
\end{eqnarray}
Another solution for $B<0$,  is   
\begin{eqnarray}
B= - \sqrt{\frac{1}{2c}n_e \delta E_F }.
\end{eqnarray}
The free energy is symmetric in $B \rightarrow -B$.

(2)  Other form of $F_0$:
\begin{eqnarray}
F_0=-\frac{1}{2 \kappa_0} B^2+\frac{\lambda_0}{4} B^4
\end{eqnarray}
 . At small $B$ of $B>B_0$ the second term is dominant and decreases with $B$, and at a large $B$ the first term is dominant and increases with $B$. Consequently, $F$ has a minimum  The first term is increasing function and the second term is decreasing function. Consequently,   $F$ has a minimum at
\begin{eqnarray}
& &\frac{ \partial F}{\partial B}= \frac{ \partial F_0}{\partial B}-\frac{1}{2}  { n_e}  \delta  E_F  \frac{1}{B} \\
& &=-\frac{1}{\kappa_0} B+\lambda_0 B^3 - \kappa \frac{n_e \delta E_F}{2} \frac{1}{B}.
\end{eqnarray}
$F$ becomes stationary at
\begin{eqnarray}
& &\lambda_0 B_0^4 -\frac{1}{\kappa_0}B_0^2-  \frac{n_e \delta E_F}{2}=0 \\
& &B_0^2=\frac{\frac{1}{\kappa_0}+ \sqrt{\frac{1}{{\kappa_0}^2}+2 \lambda_0 n_e \delta E_F}}{2 \lambda_0}
\end{eqnarray}
A stationary magnetic field shifts by an amount which depends on the product $n_e E_F$.  

$\frac{1}{\kappa_0} =0, F_0= \frac{\lambda_0}{4} B^4 $ 
\begin{eqnarray}
 B_0^2= \sqrt{\frac{  n_e \delta E_F}{2 \lambda_0}}
\end{eqnarray}
$ \lambda_0=0,F_0=\frac{1}{2 \kappa_0} B^2, \kappa_0 >0$ 
\begin{eqnarray}
 B_0^2= \kappa_0 \frac{  n_e \delta E_F}{2}
\end{eqnarray}
\subsubsection{Charged particle with spin and  separation of short range effective interaction  }
Electron has a spin $ \frac{1}{2} \hbar $ and is described by an one-particle   Hamiltonian       
\begin{align}
{H}_0=  \frac{({p_{x}}+e{ A}_x^\text{ext})^2}{  2m}+\frac{({ p_{y}}+e{ A}_y^\text{ext})^2}{  2m}+\sigma_z g \frac{eB}{2m} 
\end{align}
in a non-relativistic energy, where $g$ is a gyromagnetic ratio. A spin-orbit coupling is assumed vanishing. 

A many-body action including the  fluctuations $ A_{\mu}$ of the effective electromagnetic field in Eq.$(\ref{lagrangian})$ is  
\begin{eqnarray}
& &S=S_0+ \int d^3x [  e j_0(x) { A}_0(x) - e {\vec j}(x)\cdot{\vec { A}}(x) +j_3(x)\delta B+O({ A}_{\mu}^2)] \label{action_s} ,\\
& &S_0= \int dt  \int_{BZ}\frac{d^2 p }{  (2\pi)^2} 
 b_l^{\dagger}({\bf p}\,) 
\left[i\hbar \frac{\partial}{
 \partial t} -\tilde E_l
\right]
 b_l ({\bf  p})\nonumber
\end{eqnarray}
where 
\begin{eqnarray}
& &j_0(x)= \sum_i \psi_i^{\dagger}(x) \psi_i(x), \vec j(x)= \sum_i \psi_i^{\dagger} (\vec p+e\vec A)\psi_i(x) \\
& &j_3(x)=\sum_i \psi_i^{\dagger}(x) (\sigma_3)_{ij} \psi_j(x) \nonumber \\
& &\delta B=\partial_1 A_2-\partial_2 A_1 \nonumber
\end{eqnarray} 
and
\begin{eqnarray}
\tilde E_l^i=E_l +(\sigma_z)_{ii} g \frac{eB}{2m}
\end{eqnarray}
We do not study a spin correlation in this paper.  
\subsection{Implications of Chern-Simons term and  spectrum asymmetry}
$L_{int}$ in Eq.$(\ref{effective_lagrangian})$ and in  Eq.$( \ref{induced_lagrangian})$ has unusual properties.

(1) $L_{int}$ represent  a universal interaction  of vector potentials which neither  depend on a many-body interaction nor related with  Coulomb interaction between charges.  

 (2) $L_{int}$  in Eq.$(\ref{effective_lagrangian})$ is  bi-linear form  of $eA_{\mu}(x)$, but is reduced to  an  energy per an electron  $E_F=eA_0$ and filling fraction of electrons $\nu_e$. $L_{int}$ is reduced to a zero-th order term  in a small fluctuation of magnetic field $ \delta B$.

(3) $L_{int}$ is related with  Bohr-van Leeuwen theorem, which proves that a magnetization does not occur in a thermodynamic equilibrium limit of classical theory. Now, magnetic field breaks  time-reversal invariance and a system may not show a thermodynamic equilibrium.  This   is a feature of a system in a magnetic field.   

(4) Divergence of $L_{int}$ at $ B \rightarrow 0$  occurs in an infinite system. In a system of finite size, the Larmor radius $R$ must be smaller than the system's size. This leads a minimum magnetic field $B_0$ of the finite system.

(5)
$L_{int}$ has a connection with Slater-Pauling curves. Chern-Simons magnetic energy shifts a stationary magnetic field
due to  a long-distance fluctuation of the electrons in Landau's levels leads the energy Eq.$(\ref{induced_lagrangian}  )$ to be  proportional to   a deviation of magnetic field and makes the stationary magnetic field to shift.  A shift of the energy from this term proportional to $E_F n_e$ leads the stationary magnetic field to vary with a density of itinerant electrons. This behavior given by Eq.$(\ref{slator-pauling})$. Experimentally  a shift of the magnetic field is observed as Slater-Pauling curve.     is just the one seen in    Slater-Pauling curves. Accordingly, Slater-Pauling curves may be  an evidence   of   a energy shifts due to a stationary magnetic field  
expressed by    Chern-Simons term. 

\subsection{D=3}
In three dimensional system,  wave functions are a free  wave in a direction of the magnetic field. The calculation is straightforward, and  we show the result here. The action is equivalent to the 2 spatial dimensions,  and is  given as
 \begin{eqnarray}
& &S_{eff}=S_\text{CS}+S_{high}, \\
& & S_\text{CS}=e^2\frac{\sigma_{xy}^{(4)}}{ 2} \int d^4 x 
 \epsilon^{\mu\nu\rho} { A}_{\mu}
 \partial_{\nu}{ A}_{\rho}, \label{chern-simon} \\
& &\sigma_{xy}^{(4)}=\frac{1}{ 2\pi} \nu^{(4)},\ \nu^{(4)}={ 2\pi} \nu
= \frac{2\pi n_e}{  eB },
\end{eqnarray}
where $\mu,\nu,\rho$ are in a space perpendicular to the magnetic field, $(0,1,2)$. Other quantities are the same as before.  
\subsubsection{Implications  of $ S_{\text{CS}}$ in radiative  scatterings  }
Eq.$(\ref{chern-simon})$  is a bi-linear  form of vector potentials, and   represents  
  a long-range correlation, which  is unrelated with shrt-range correlations.
Short range correlations modify an energy, mass, magnetic moment, and other parameters 
of a system and renormalized quantities of the system.  Long-range correlations, however,  barely affect in short distance and do not renormalize these parameters.
 
 
We write a typical length of  physical system $l_{ph}$.  If  de Broglie wave length is an only typical length  $l_{ph}=l_w$ in a physical system,  a transition  probability and other physical quantities are determined  by  short range correlations.  An energy of the system and  scattering cross section are  governed by the 
length,   $l_w=\frac{h}{p}$, in fact. This length  for $p \geq 1$ keV, is order of  atomic sizes whereas   the magnetic length  $R_{B}=\frac{mv}{eB}$    for $B= 1$ Tesla and $1$ eV    
 is   { $R_B\approx   10^{-8}$} M . The magnetic length   is much longer than the atomic distance, and its fluctuations do not contribute to an  amplitude   in   short distance regions.  
Interactions  in short distance regions   are  not modified by 
   the     topological interaction Eq.$(\ref{chern-simon})$.


 $R_B$ is a new scale that contribute to many-body wave functions as well.  
 Effects specific  to these waves   appear  in a system of  length $l_{ph}$ of satisfying $l_{ph} \gg  R_{B}$ and are described by  the action $ S_\text{CS} $, but
  these  disappear  in  the short distance of $l_{ph}  \ll  R_{B}$.  
 $ S_\text{CS} $  is the action  of the lowest dimension      reveals  the phenomena of   physical length $l_{ph} \gg R_{B}$. 
        A property at a long distance     is similar to the one of the  EPR correlation \cite{EPR}.

\subsection{$B^n$ terms from short range interaction}
Energies of the bound electrons are also modified by a magnetic filed. This effects come from short distance correlations inside atom.   From Fig.(1),  this region belongs to weak field region. A  correction is computed by an energy shift of the atomic levels, Zeeman shift,
\begin{eqnarray}
\delta E= B \mu_l
\end{eqnarray}
and higher order terms,
\begin{eqnarray}
\delta E=B^2, B^n, n>3
\end{eqnarray}

  The   energy from the short distance fluctuations   behaves as $B^n,:n=0,1,2,\cdots$.


The sum of the energy proportional to $\frac{1}{B}$ Eq.$(\ref{induced_lagrangian})$  and
  another term of  $B^n, n>1$ has a minimum at one $B$.  Accordingly  a spontaneous magnetization occurs.

\subsection{Connection with an orbital magnetic moment?}
     
The effective Lagrangian Eq.$(\ref{effective-magnetic field lagrangian})$  could be understood as a manifestation of an orbital 
magnetic moment of electrons \cite{orbital}. For electron bound states,   the orbital angular momentum plays a similar role as the spin, and couples with a magnetic field. However, this is not a case for itinerant electrons, where an orbital angular momentum is not well defined. In fact, the present result does not rely on linear response approximation, and is not proportional to the magnetic field.
 It is hard to assign the orbital magnetic moment from the present result.  
\section{Summary}
The magnetic properties of the itinerant magnetism is studied with the  effective action of  electromagnetic vector potentials.  The action   is found to include  a topological Chern-Simons term in the strong field region in Fig.1. The Chern-Simons term has the lowest scale dimension among gauge invariant quantites and determines a long distance electromagnetic properties of physical systems.  The system of a finite electron density and a magnetic field is described by a constant $A_0$ and  $B$ which is expressed by linear   
 vector potentials. Unique properties of the present system nonexisting  in the vacuum are understood from  Chern-Simons term.  For its computation,  the creation and annihilation operators of  electrons  defined with energy eigenstates  and a propagator  defined in von Neumann representation of Landau levels are used. These are convenient to study  universal  properties of itinerant electrons in a magnetic filed 
and is computed    based on  $\frac{1}{B}$ expansion. 

The $1/B$ expansion  is valid  in large systems even at a finite or  weak magnetic field. Chern-Simons term in the effective Lagrangian is responsible to  Hall effect  in a system under inject current, and to  system's energy in the absence of external current.    Because the    energy 
  has a dependence on the magnetic field, it   affects to system's magnetic property. One characteristic feature arises in
a stationary magnetic field that varies  universally when the electron density is varied as in  Eq.$(\ref{slator-pauling})$. Experimentally  a shift of the magnetic field is observed as Slater-Pauling curve.   The universal behavior of
Slater-Pauling curve   is due to the topological property of Chern-Simons term, which is not connected with the spin.   

\subsection*{Acknowledgments}
 This work was partially supported by a 
Grant-in-Aid for Scientific Research ( Grant No. 24340043). The authors thank Drs. Amitsuka,   Goryo and Matsuyama for useful discussions.

\appendix 
\section{ Landau levels of two-band equation}
\begin{eqnarray}
& &( i \frac{\partial}{\partial t} -m)\phi=\bs \sigma \cdot( \bs p+e \bs A)\xi ,   \\
& & (i \frac{\partial}{\partial t} +m) \xi =\bs \sigma \cdot( \bs p+e \bs A)\phi \nonumber , 
\end{eqnarray}
The stationary solutions $\phi=e^{i E t} \phi, \xi=e^{i E t} \xi $
\begin{eqnarray}
& &(E  -m)\phi=\bs \sigma \cdot( \bs p+e \bs A)\xi ,   \\
& & (E +m) \xi =\bs \sigma \cdot( \bs p+e \bs A)\phi  \nonumber , 
\end{eqnarray}
\begin{eqnarray}
& &(E^2  -m^2)\phi=[  {\bs p}^2 +e^2 {\bs A}^2 +2 e \bs A \cdot \bs p   +e \bs \sigma \cdot \bs H ] \phi \\
& &E=\pm\sqrt{m^2+p_z^2+(2l+1+\sigma) eB  } \nonumber
\end{eqnarray}
In the non-relativistic region, the energy is expressed as
\begin{eqnarray}
E=\pm [ m+ \frac{p_z^2+(2l+1+\sigma) eB }{2m}+higher orders]
\end{eqnarray}

\end{document}